\documentclass[9pt,twocolumn,twoside]{osajnl}
\journal{ol} 
\setboolean{shortarticle}{true}

\renewcommand{\journalshorttype}{}
\renewcommand{\journallongtype}{}
\renewcommand{\journalname}{}
\usepackage[textwidth=1.0 cm]{todonotes}
\usepackage{braket}
\usepackage{subfigure}
\usepackage{comment}
\usepackage{ulem}
\newcommand{\two}[1]{{\color{red} #1}}
\newcommand{\one}[1]{{\color{blue} #1}}

\definecolor{color2b}{rgb}{0.2,0.2,0.7} 
\begin{document}
\title{Pure-Quartic Optical Shock Waves}
\author[1]{Sanjida Yeasmin}
\author[2,3]{Sathyanarayanan Chandramouli}
\author[1]{Ziad H. Musslimani}
\author[4]{Andrea Blanco-Redondo}
\affil[1]{Department of Mathematics, Florida State University, Tallahassee, FL 32306, USA}
\affil[2]{Department of Mathematics and Statistics, University of Massachusetts at Amherst, Massachusetts, 01003, USA}
\affil[3]{Okinawa Institute of Science and Technology, Onna, Okinawa 904-0497, Japan}
\affil[4]{CREOL, The College of Optics and Photonics, University of Central Florida, Orlando, FL 32816, USA}
\begin{abstract}
In this paper, we study the emergence and dynamics of pure-quartic optical dispersive shock waves governed by the nonlinear Schr\"odinger (NLS) equation with self-defocusing Kerr nonlinearity and fourth-order dispersion. The corresponding dispersionless hydrodynamic system reveals a distinct mechanism for wave breaking, driven by the nonlinear self-steepening of the hydrodynamic velocity. We illustrate this mechanism through optical dam-break configurations described by Riemann problems and show that, in contrast to the classical quadratic NLS equation, wave breaking occurs prior to wave splitting even in the small amplitude regime. Numerical simulations of the pure-quartic NLS (PQNLS) equation and its hydrodynamic approximation confirm these qualitatively distinct dynamical regimes.
\end{abstract}
\setboolean{displaycopyright}{true}
\setlength{\marginparwidth}{2cm}
\maketitle
\begin{figure}[hbt!]
    \centering
\includegraphics[width=1\linewidth]{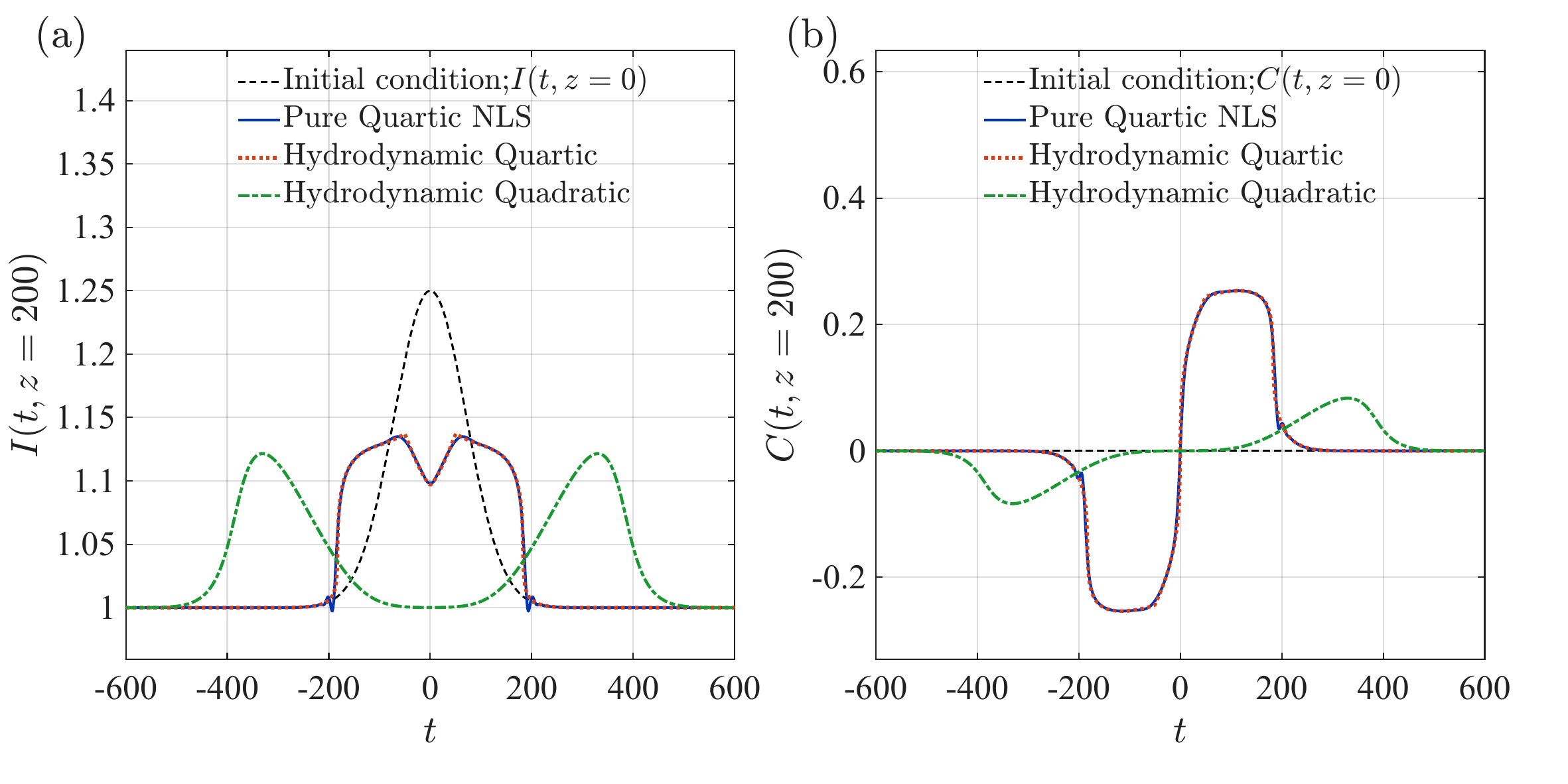}    
\caption{Propagation of a Gaussian input beam with intensity {$I(t,z=0)=1+0.25\exp(-0.001t^2)$}, and zero initial hydrodynamic velocity, $C(t,z=0)=0$. 
The blue solid, red dashed, and green dashed curves correspond to the pure-quartic NLS equation [Eq.~(\ref{PQNLS})], the corresponding hydrodynamic system [Eqns.~(\ref{DH-1}) and (\ref{DH-2})], and the hydrodynamic equations in the quadratic NLS limit, respectively, while the black dashed curve denotes the initial condition. (a) Intensity. (b) Hydrodynamic velocity.}
\label{DH-1-2-sim}
\end{figure}
\begin{figure}[hbt!]
    \centering    \includegraphics[width=1\linewidth]{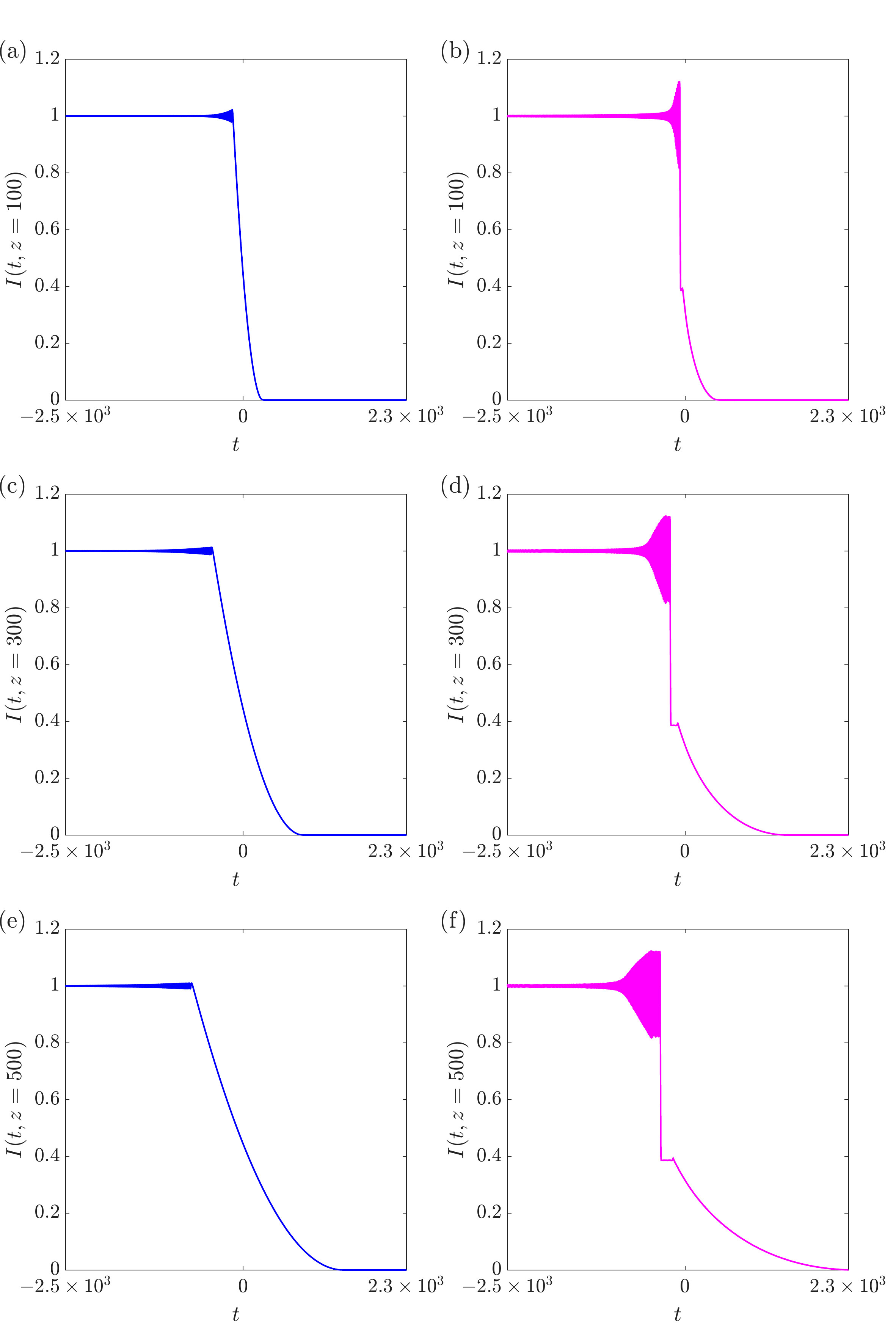}
   \caption{The dam break problem (\eqref{DAM-BREAK}) for the quadratic (left panel) and quartic (right panel). Snapshots of the intensity at several propagation distances, shown from top to bottom. Panels (a), (c), and (e) show the quadratic case, while panels (b), (d), and (f) show the pure-quartic NLS. The initial dam-break profile develops into a self-similar rarefaction in the quadratic case, whereas the pure-quartic case produces counterpropagating rarefaction (right) and oscillatory (left) waveforms [see the text for a description].}
    \label{Dam-breaking}
\end{figure}
\begin{figure}
    \centering
    \includegraphics[width=1\linewidth]{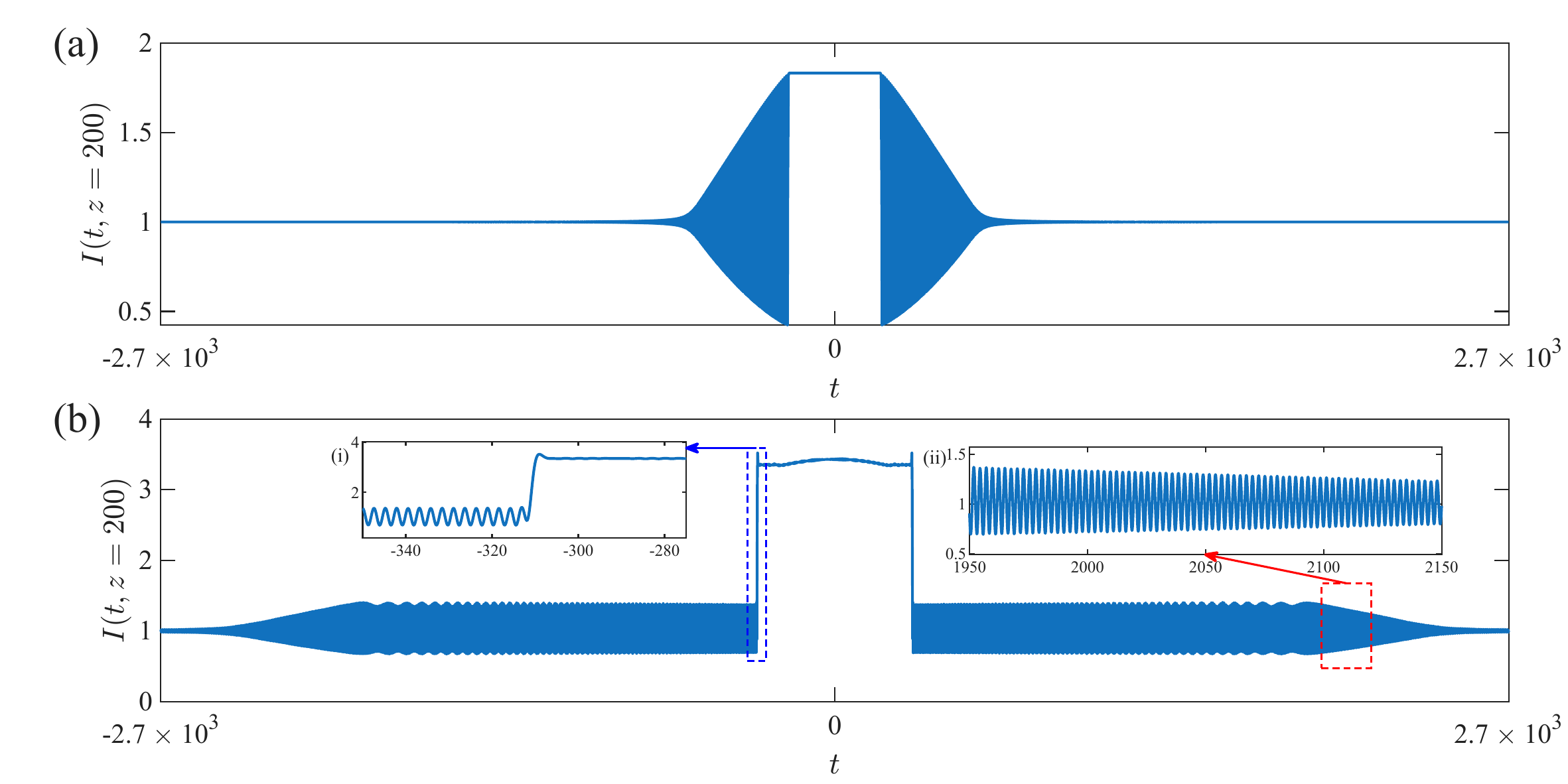}
   \caption{The piston Riemann problem for the quadratic (a) and quartic (b) NLS equations. (a) Intensity snapshot for the quadratic NLS at $z=200$, depicting two counterpropagating DSWs. (b) Intensity snapshot for the quartic NLS at $z=200$, displaying counterpropagating tDSWs separated by an expanding plateau. The insets, from left to right, show the heteroclinic connection between the periodic-wave and plateau regions and the partial-DSW structure, within the body of the tDSW respectively. }
   \label{Figure-Piston-Riemann}
\end{figure}

Dispersive shock waves (DSWs) are nonlinear coherent structures that emerge when wave breaking is regularized by dispersion \cite{el2016dispersive,el2005resolution,gurevich1973nonstationary}. Unlike classical shock waves, where discontinuities in physical quantities, such as density, pressure, and velocity are regularized by viscous effects, DSWs consist of expanding regions of rapidly oscillating nonlinear wave trains that connect two distinct uniform-intensity states. In recent years, 
DSWs have become an active area of research across a wide range of scientific disciplines, including nonlinear optics \cite{wan2007dispersive,bendahmane2022piston,jia2007dispersive,conti2009observation},  fluid mechanics \cite{trillo2016observation,maiden2016observation,el2012transformation,el2006unsteady}, and ultracold atomic gases \cite{hoefer2006dispersive,chang2008formation,kevrekidis2008emergent}, among others.\\
DSWs have been experimentally realized in a variety of nonlinear photonic platforms, including optical fibers \cite{trillo2017wave}, photorefractive crystals \cite{wan2007dispersive}, and waveguide arrays \cite{jia2007dispersive}. These nonlinear phenomena, over a broad swathe of instances, can be analytically described using the nonlinear Schr\"odinger (NLS)  equation with self-defocusing nonlinearity and quadratic dispersion.\\
The role of competing higher-order dispersion in DSW generation has received significant attention in recent studies, most notably in connection with the resonant radiation emitted during temporal pulse propagation in optical fibers \cite{bakholdin2004non,conforti2014resonant} and during beam propagation in nematic liquid crystals \cite{el2016radiating}. This has important implictaions in supercontinuum generation \cite{hooper2011coherent,xu2012supercontinuum,conforti2013dispersive} and analogous phenomena in fluid mechanics \cite{sprenger2017shock,baqer2025shallow} and in ultracold atomic gases with ``negative effective'' mass \cite{mossman2020stability}. In contrast to such competing-dispersion settings, pure-quartic dispersion constitutes a distinct higher-order dispersive regime in which conventional quadratic dispersion is absent (or negligible), rather than merely supplemented by higher-order corrections. A fundamental advance in this direction was the experimental discovery of pure-quartic solitons (PQS)\cite{blanco2016pure}, and the subsequent realization of the PQS laser \cite{runge2020pure}, which opened new avenues for investigating nonlinear pulse propagation in regimes dominated by fourth-order and other even high-orders of dispersion \cite{tam2019stationary,de2021pure, runge2021infinite, liu2022dynamic, qian2022dissipative, de2023even, sardelis2024pure}. In the context of optical DSWs, the replacement of conventional quadratic dispersion by quartic dispersion can fundamentally alter their formation and evolution by modifying the underlying dispersive hydrodynamics of the optical fluid, whose density and hydrodynamic velocity are represented by the optical intensity and chirp, respectively.\\
In this paper, we investigate the formation and physical mechanisms underlying optical dispersive shock waves in the PQNLS with self-defocusing Kerr nonlinearity and fourth-order dispersion. We examine how the absence of conventional quadratic dispersion fundamentally alters the dispersionless hydrodynamic behavior, leading to wave phenomena that differ from those of the classical NLS equation. Using optical dam-break configurations and Riemann problems as prototypical settings, we analyze the development of steep intensity and chirp profiles and identify the mechanisms responsible for nonlinear wave breaking and wave splitting. The hydrodynamic formulation provides a direct framework for describing these processes and reveals the role of nonlinear advection in the steepening of the hydrodynamic velocity. Comparisons with the quadratic NLS case demonstrate a qualitatively different ordering of wave splitting and breaking, underscoring the distinctive role of quartic dispersion in shaping the structure and propagation of optical dispersive shock waves. Numerical simulations of both the full wave equation and its dispersionless hydrodynamic approximation substantiate the analytical picture and clarify the range of validity of the hydrodynamic description. \\
We begin by considering the NLS equation governing the evolution of the complex optical envelope, $\Psi(t,z)$ in the presence of fourth-order dispersion and Kerr nonlinearity:
\begin{equation}
    i \frac{\partial \Psi}{\partial Z} = \frac{\beta_4}{24} \frac{\partial^4 \Psi}{\partial T^4} + \gamma_{\rm eff} |\Psi|^2 \Psi \;.
    \label{eq:dimensional}
\end{equation}
Here, $T$ denotes the retarded time, $Z$ is the propagation distance, $\beta_4$ is the fourth-order dispersion coefficient, and $\gamma_{{\rm eff}}$ characterizes the effective nonlinear strength. To obtain a dimensionless formulation, we introduce the characteristic temporal and spatial scales associated with fourth-order dispersion, together with a characteristic field amplitude. Specifically, we define
$t = T/T_0,~ z = Z/L_{FOD},$ and $\psi(z, t) = \Psi(Z, T) / A$,
where the fourth-order-dispersion length is denoted by $L_{FOD}$ \cite{blanco2016pure}, and
\begin{equation}
T_0 = \left( \frac{\beta_4 L_{\text{FOD}}}{24} \right)^{\frac{1}{4}} , \quad 
A = \left( \frac{1}{\gamma L_{FOD}} \right)^{\frac{1}{2}} \;.
\end{equation}
With these scalings, the dimensional equation can be reduced to the dimensionless temporal PQNLS equation 
\begin{equation}
    i \frac{\partial \psi}{\partial z} = \frac{\partial^4 \psi}{\partial t^4} + |\psi|^2 \psi \;,
    \label{PQNLS}
\end{equation}
where $t$ and $z$ denotes the dimensionless temporal coordinate and propagation distance, respectively.
An optical fluid interpretation of the PQNLS equation is obtained through the Madelung transformation
\begin{equation}
 \psi(t,z) = \sqrt{ I (t,z)}\exp(i\phi(t,z)) \;,
 \label{Madelung}
\end{equation}
where $I$ and $\phi$ denote the optical intensity and phase, respectively. 
Substituting \eqref{Madelung} into \eqref{PQNLS} and neglecting higher-order derivative terms, justified for a sufficiently slowly varying optical field and for propagation distances prior to the onset of gradient catastrophe, yields the following approximate compressible Euler-type (dispersionless) hydrodynamic equations:
\begin{equation}
    \label{DH-1}
    \frac{\partial I}{\partial z}  + 4 \frac{\partial}{\partial t}  ( I C^3 ) \approx 0 \;,
\end{equation}
\begin{equation}
    \label{DH-2}
    \frac{\partial C}{\partial z} +  \frac{\partial}{\partial t} ( I+ C^4 ) \approx 0 \;.
\end{equation}
Here, the optical chirp, $C \equiv \phi_t$ denotes the hydrodynamic velocity.
Equations (\ref{DH-1}) and (\ref{DH-2}) constitute a hyperbolic system of partial differential equations \cite{lax2006hyperbolic,whitham2011linear} and, as such, can exhibit wave breaking and shock formation -- features that are not directly apparent in the PQNLS equation itself and become evident through the Madelung transformation. 
To illustrate these behaviors, we present in Fig.~\ref{DH-1-2-sim} numerical simulations of the hydrodynamic equations with a small-amplitude Gaussian input intensity and zero initial phase ($\phi=0$). 
A preliminary insight into the dynamics of the intensity and chirp can be obtained by linearizing the hydrodynamic equations about the uniform background state $I= 1$ and $C=0$. Writing $I(t,z)= 1 +\tilde{I}(t,z)$ and $C(t,z)= \tilde{C}(t,z)$, and retaining only leading order terms, we obtain the evolution equations $\tilde{I}_z\approx 0$ and $\tilde{C}_z\approx -\tilde{I}_t.$ The solutions of these equations are $\widetilde{I}(t,z)\approx \widetilde{I}(t,0)$ and $\widetilde{C}(t,z)\approx -z\partial_t\widetilde{I}(t,0)$, indicating a nearly stationary intensity profile accompanied by a double-lobed chirp pinned at $t=0$, with lobe amplitudes that grow linearly with propagation distance. To gain physical insight into the wave-breaking mechanism, we rewrite the hydrodynamic equations in 
non-conservative form, thereby highlighting the role of nonlinear advection in the beam propagation dynamics. The intensity equation becomes
$ I_z  +  4 C^3~I_t  \approx -  12 I C^2 C_t,$ 
while the hydrodynamic velocity satisfies
$C_z  +  4 C^3 C_t  \approx - I_t$.
An intuitive interpretation can be obtained by extending the linear analysis presented above and assuming that the intensity profile remains nearly stationary. In this regime, the intensity gradient $I_t$ 
acts as a driving force that generates a nonzero hydrodynamic velocity through the second equation. Once generated, this velocity is advected according to $C_z  +  4 C^3 C_t $, with a characteristic advection speed proportional to $C^3$. Consequently, regions with larger local velocities, whose generation is explicable through the linear analysis above, propagate faster than those with smaller ones, resulting in progressive self-steepening and eventual wave breaking.
For comparison, let us now consider the NLS equation with quadratic dispersion. In this case, the equations governing the intensity and hydrodynamic velocity take the form
$ I_z  +  C~I_t  + I C_t\approx 0$, 
and
$C_z  +  2 C C_t + I_t\approx 0$.
A linearization of this hydrodynamic system reveals a qualitatively different mechanism governing the initial wave dynamics. Specifically, the linearized equations form the coupled system 
$\tilde{I}_{z}+2\tilde{C}_t\approx 0$ and $\tilde{C}_z+\tilde{I}_t\approx 0$. Combining these equations yields the two-way wave equation 
$\tilde{I}_{zz}-2 \tilde{I}_{tt} \approx 0$  , which describes counterpropagating intensity and velocity perturbations. 
Thus, in contrast to the quartic NLS case, the initially localized small-amplitude profile is expected to split into two counterpropagating components, while the linearized velocity dynamics do not exhibit sustained amplitude growth. This suggests that wave breaking occurs only after wave splitting. See \cite{isoard2019wave} for a detailed discussion of the corresponding dynamics in the quadratic NLS case.\\
The dynamics of the PQNLS and quadratic NLS equations, together with their respective hydrodynamic approximations, are summarized in Fig.~\ref{DH-1-2-sim}. Equation~(\ref{PQNLS}) is solved numerically using the fourth-order exponential time-differencing Runge--Kutta (ETDRK4) spectral method \cite{kassam2005fourth}, with the initial intensity profile shown in Fig.~\ref{DH-1-2-sim}. The corresponding dispersionless hydrodynamic system is integrated using the Richtmyer two-step Lax--Wendroff scheme \cite{lax1960systems,richtmyer1967difference}. The same numerical methods are employed for the pure quadratic NLS equation and its corresponding dispersionless hydrodynamic approximation. In addition to verifying the fidelity of the hydrodynamic approximations in this asymptotic regime, the results highlight a clear distinction between wave splitting and wave breaking: the PQNLS dynamics exhibit wave breaking before splitting, whereas the quadratic NLS dynamics exhibit splitting before wave breaking. A more complete understanding of the wave-breaking process in the quartic NLS case requires a deeper analysis of the properties of the corresponding PQNLS hydrodynamic system, which is deferred to a separate study.
Having examined the wave-breaking behavior of the PQNLS hydrodynamic system, which develops over relatively short propagation distances, we now turn our attention to how this nonlinear steepening is regularized by dispersion, leading to the formation of a coherent nonlinear structure. To this end, we consider a configuration known as a dam break. Rooted in fluid mechanics, a dam break \cite{Stoker1957,Ritter1892} occurs when a barrier separating two regions of fluid with different depths is suddenly removed. The resulting flow generates nonlinear waves driven by the associated hydrodynamic pressure gradient. Mathematically, a dam-break scenario can be formulated as a specific class of Riemann problem, namely, an initial-value problem characterized by step-like initial data in the intensity and zero initial velocity. \\
To do so, we consider the optical analogue of the dam-break configuration, consisting of two well-separated uniform-intensity states (one of which is the zero state) connected by a smoothened intensity transition. This setting, in turn, defines a Riemann-type initial-value problem for the PQNLS equation. We numerically solve Eq.~(\ref{PQNLS}) using ETDRK4, subject to the initial intensity and phase [see Eq.~(\ref{Madelung})]
\begin{equation}
\label{DAM-BREAK}
I(t,z=0) = \frac{1}{2} \left( 1 - \tanh (t-t_0) \right) \;,\;\;\; C(t,z=0)=0 \;.
\end{equation}
To facilitate the use of the Fourier spectral method, we employ a box-type initial profile, with the smoothed intensity transition in \eqref{DAM-BREAK} corresponding to one half of the box. The numerical results at several propagation distances are shown in Fig.~\ref{Dam-breaking}. For the PQNLS equation, shown in panels \one{(b), (d), and (f)}, the optical dam-break profile evolves into bidirectionally propagating waves. The right-propagating waveform spreads in a self-similar fashion and is referred to as a rarefaction wave \cite{el2016dispersive}. The left-propagating waveform is composite, comprising a kink and an oscillatory structure that propagates upstream at a slightly higher speed. The oscillatory structure is referred to as a partial DSW \cite{sprenger2017shock,el2006unsteady}, since its modulations do not extend to the solitonic limit, but instead terminate at a uniform periodic state. Such composite waveforms do not have an analog in the quadratic NLS, where dam break results in a single rarefaction waveform \cite{el2016dispersive,el1995decay}, and are a genuine novelty of the quartic dispersion (see also related works \cite{mohapatra2026dam,kamchatnov2012undular}). This fundamentally different dam-break dynamics could have implications in high-energy coherent spectral shaping, a crucial application of optical wave breaking, provided that quartic-shock formation can produce broader spectra and higher-energy compressible pulses before other detrimental effects appear, which remains an element of study.
Another fundamental paradigm for shock-wave generation, originating in fluid mechanics, is the piston problem of gas dynamics (see, e.g., \cite{coulson1977waves}), in which shock waves are generated by the compressive action of a piston on a gas initially at rest. Analogous problems have been extended to a variety of physical settings, including optics \cite{bendahmane2022piston}, superfluids \cite{hoefer2008piston}, and ferromagnetics \cite{hu2022spin}, where the viscous shocks of gas dynamics are replaced by their dispersive counterparts. The piston Riemann problem considers the simultaneous action of two compressive pistons \cite{bendahmane2022piston}. Mathematically, it may be viewed as the complement of the dam-break problem; consisting of a constant intensity profile with a smoothened transition in the velocity that facilitates compression (c.f. \eqref{Madelung})
\begin{equation}
\label{Velocity-Riemann}
I(t,z=0) = 1\;,\;\;\;\;\; C(t,z=0) =- \tanh(t-t_0)\,\;.
\end{equation}
Numerically, the equations are integrated using the ETDRK4 Fourier-spectral method, with the initial condition in \eqref{Velocity-Riemann} suitably mollified to enforce vanishing boundary values on the computational domain. The numerical results at the propagation distance $z=200$ are shown for the quadratic and quartic NLS in Fig.~\ref{Figure-Piston-Riemann}(a) and (b), respectively. 
For the quadratic NLS, the piston Riemann problem generates two counterpropagating DSWs of equal strength, separated by an expanding plateau \cite{el2016dispersive}. In contrast, for the same jump parameters, the quartic NLS gives rise to a pair of counterpropagating waveforms reminiscent of traveling dispersive shock waves (tDSWs) \cite{sprenger2017shock}. Such structures are characteristic of optical \cite{el2016radiating}, superfluidic \cite{yang2026dispersive}, and fluid-mechanical \cite{sprenger2017shock,baqer2025shallow} systems with higher-order, nonconvex dispersion. Structurally, the tDSW differs significantly from a conventional DSW, consisting of a partial DSW that transitions into a uniform periodic state, followed by a rapid transition to a uniform-intensity plateau (see Fig.~\ref{Figure-Piston-Riemann}(b) and the corresponding insets). In this way, the pure-quartic hydrodynamics within the piston-Riemann problem resemble a nonlinear traveling information carrier, rather than an expanding shock fan as in the quadratic case, which opens the door to future studies of information transport by the shock structure.



\section*{Disclosures}
 The authors declare no conflicts of interest.
 
\section*{Data availability}
No data were generated or analyzed in the presented research.

\bibliography{references}

@article{el1995decay,
  title={Decay of an initial discontinuity in the defocusing NLS hydrodynamics},
  author={El, Gennady A and Geogjaev, VV and Gurevich, AV and Krylov, AL},
  journal={Physica D: Nonlinear Phenomena},
  volume={87},
  number={1-4},
  pages={186--192},
  year={1995},
  publisher={Elsevier}
}

@article{coulson1977waves,
  title={Waves: A mathematical approach to the common types of wave motion},
  author={Coulson, Charles Alfred and Jeffrey, Alan},
  journal={(No Title)},
  year={1977}
}

@article{bendahmane2022piston,
  title={The piston Riemann problem in a photon superfluid},
  author={Bendahmane, Abdelkrim and Xu, Gang and Conforti, Matteo and Kudlinski, Alexandre and Mussot, Arnaud and Trillo, Stefano},
  journal={Nature Communications},
  volume={13},
  number={1},
  pages={3137},
  year={2022},
  publisher={Nature Publishing Group UK London}
}

@article{mohapatra2026dam,
  title={Dam breaks in the discrete nonlinear Schr{\"o}dinger equation},
  author={Mohapatra, Shrohan and Kevrekidis, Panayotis G and Yang, Su and Chandramouli, Sathyanarayanan},
  journal={Physica D: Nonlinear Phenomena},
  pages={135279},
  year={2026},
  publisher={Elsevier}
}

@article{kamchatnov2012undular,
  title={Undular bore theory for the Gardner equation},
  author={Kamchatnov, AM and Kuo, Y-H and Lin, T-C and Horng, T-L and Gou, S-C and Clift, Richard and El, GA and Grimshaw, Roger HJ},
  journal={Physical Review E--Statistical, Nonlinear, and Soft Matter Physics},
  volume={86},
  number={3},
  pages={036605},
  year={2012},
  publisher={APS}
}

@article{isoard2019wave,
  title={Wave breaking and formation of dispersive shock waves in a defocusing nonlinear optical material},
  author={Isoard, M and Kamchatnov, AM and Pavloff, N},
  journal={Physical Review A},
  volume={99},
  number={5},
  pages={053819},
  year={2019},
  publisher={APS}
}

@article{hoefer2008piston,
  title={Piston dispersive shock wave problem},
  author={Hoefer, Mark A and Ablowitz, Mark J and Engels, Peter},
  journal={Physical Review Letters},
  volume={100},
  number={8},
  pages={084504},
  year={2008},
  publisher={APS}
}

@article{el2016dispersive,

  title={Dispersive shock waves and modulation theory},

  author={El, GA and Hoefer, MA},

  journal={Physica D: Nonlinear Phenomena},

  volume={333},

  pages={11--65},

  year={2016},

  publisher={Elsevier}

}

@article{gurevich1973nonstationary,
  title={Nonstationary structure of a collisionless shock wave},
  author={Gurevich, AV and Pitaevskii, LP},
  journal={Zhurnal Eksperimentalnoi i Teoreticheskoi Fiziki},
  volume={65},
  pages={590--604},
  year={1973}
}

@article{conti2009observation,
  title={Observation of a gradient catastrophe generating solitons},
  author={Conti, Claudio and Fratalocchi, Andrea and Peccianti, Marco and Ruocco, Giancarlo and Trillo, Stefano},
  journal={Physical Review Letters},
  volume={102},
  number={8},
  pages={083902},
  year={2009},
  publisher={APS}
}

@article{hu2022spin,
  title={Spin-piston problem for a ferromagnetic thin film: Shock waves and solitons},
  author={Hu, Mingyu and Iacocca, Ezio and Hoefer, Mark A},
  journal={Physical Review B},
  volume={105},
  number={10},
  pages={104419},
  year={2022},
  publisher={APS}
}

@article{jia2007dispersive,
  title={Dispersive shock waves in nonlinear arrays},
  author={Jia, Shu and Wan, Wenjie and Fleischer, Jason W},
  journal={Physical Review Letters},
  volume={99},
  number={22},
  pages={223901},
  year={2007},
  publisher={APS}
}

@article{el2005resolution,
  title={Resolution of a shock in hyperbolic systems modified by weak dispersion},
  author={El, Gennady A},
  journal={Chaos: An Interdisciplinary Journal of Nonlinear Science},
  volume={15},
  number={3},
  pages={037103},
  year={2005},
  publisher={American Institute of Physics}
}

@article{maiden2016observation,
  title     = {Observation of Dispersive Shock Waves, Solitons, and Their Interactions in Viscous Fluid Conduits},
  author    = {Maiden, Michelle D. and Lowman, Nicholas K. and Anderson, Dalton V. and Schubert, Marika E. and Hoefer, Mark A.},
  journal   = {Physical Review Letters},
  volume    = {116},
  number    = {17},
  pages     = {174501},
  year      = {2016},
  month     = {April},
  url       = {https://link.aps.org/doi/10.1103/PhysRevLett.116.174501},
  doi       = {10.1103/PhysRevLett.116.174501}
}

@article{el2012transformation,
  title     = {Transformation of a Shoaling Undular Bore},
  author    = {El, G. A. and Grimshaw, Roger H. J. and Tiong, Wei K.},
  journal   = {Journal of Fluid Mechanics},
  volume    = {709},
  pages     = {371--395},
  year      = {2012},
  publisher = {Cambridge University Press}
}

@article{trillo2016observation,
  title     = {Experimental Observation and Theoretical Description of Multisoliton Fission in Shallow Water},
  author    = {Trillo, S. and Deng, G. and Biondini, G. and Klein, M. and Clauss, G. F. and Chabchoub, A. and Onorato, M.},
  journal   = {Physical Review Letters},
  volume    = {117},
  number    = {14},
  pages     = {144102},
  year      = {2016},
  month     = {Sep},
  url       = {https://link.aps.org/doi/10.1103/PhysRevLett.117.144102},
  doi       = {10.1103/PhysRevLett.117.144102}
}

@article{yang2026dispersive,
  title={Dispersive shock waves in periodic lattices},
  author={Yang, Su and Chandramouli, Sathyanarayanan and Kevrekidis, Panayotis G},
  journal={Physical Review E},
  volume={113},
  number={4},
  pages={044204},
  year={2026},
  publisher={APS}
}

@article{wan2007dispersive,
  title     = {Dispersive Superfluid-like Shock Waves in Nonlinear Optics},
  author    = {Wan, Wenjie and Jia, Shu and Fleischer, Jason W.},
  journal   = {Nature Physics},
  volume    = {3},
  number    = {1},
  pages     = {46--51},
  year      = {2007}
}

@book{richtmyer1967difference,
  title     = {Difference Methods for Initial-Value Problems},
  author    = {Richtmyer, Robert D. and Morton, K. W.},
  edition   = {2},
  publisher = {Interscience Publishers, John Wiley \& Sons},
  address   = {New York},
  year      = {1967}
}

@article{lax1960systems,
  title={Systems of conservation laws},
  author={Lax, Peter D. and Wendroff, Burton},
  journal={Communications on Pure and Applied Mathematics},
  volume={13},
  number={2},
  pages={217--237},
  year={1960},
  publisher={Wiley}
}

@article{kassam2005fourth,
  title={Fourth-order time-stepping for stiff PDEs},
  author={Kassam, Aly-Khan and Trefethen, Lloyd N},
  journal={SIAM Journal on Scientific Computing},
  volume={26},
  number={4},
  pages={1214--1233},
  year={2005},
  publisher={SIAM}
}

@book{Stoker1957,
  author    = {Stoker, James Johnston},
  title     = {Water Waves: The Mathematical Theory with Applications},
  series    = {Pure and Applied Mathematics},
  volume    = {4},
  publisher = {Interscience Publishers},
  address   = {New York},
  year      = {1957}
}

@article{Ritter1892,
  author  = {Ritter, August},
  title   = {Die Fortpflanzung von Wasserwellen},
  journal = {Zeitschrift des Vereines Deutscher Ingenieure},
  volume  = {36},
  number  = {33},
  pages   = {947--954},
  year    = {1892}
}

@article{bakholdin2004non,
  title={Non-dissipative discontinuities in continuum mechanics},
  author={Bakholdin, IB},
  journal={M: Physmathlit},
  year={2004}
}

@book{whitham2011linear,
  title={Linear and nonlinear waves},
  author={Whitham, Gerald Beresford},
  year={2011},
  publisher={John Wiley \& Sons}
}

@article{blanco2016pure,
  title={Pure-quartic solitons},
  author={Blanco-Redondo, Andrea and De Sterke, C Martijn and Sipe, John E and Krauss, Thomas F and Eggleton, Benjamin J and Husko, Chad},
  journal={Nature Communications},
  volume={7},
  number={1},
  pages={10427},
  year={2016},
  publisher={Nature Publishing Group UK London}
}

@book{lax2006hyperbolic,
  title     = {Hyperbolic Partial Differential Equations},
  author    = {Lax, Peter D.},
  publisher = {American Mathematical Society},
  year      = {2006}
}

@article{de2021pure,
  title={Pure-quartic solitons and their generalizations—Theory and experiments},
  author={de Sterke, C Martijn and Runge, Antoine FJ and Hudson, Darren D and Blanco-Redondo, Andrea},
  journal={Apl Photonics},
  volume={6},
  number={9},
  year={2021},
  publisher={AIP Publishing}
}

@article{runge2020pure,
  title={The pure-quartic soliton laser},
  author={Runge, Antoine FJ and Hudson, Darren D and Tam, Kevin KK and de Sterke, C Martijn and Blanco-Redondo, Andrea},
  journal={Nature Photonics},
  volume={14},
  number={8},
  pages={492--497},
  year={2020},
  publisher={Nature Publishing Group UK London}
}

@article{tam2019stationary,
  title={Stationary and dynamical properties of pure-quartic solitons},
  author={Tam, Kevin KK and Alexander, Tristram J and Blanco-Redondo, Andrea and Martijn de Sterke, C},
  journal={Optics Letters},
  volume={44},
  number={13},
  pages={3306--3309},
  year={2019},
  publisher={Optical Society of America}
}

@article{de2023even,
  title={Even-order dispersion solitons: A pedagogical note},
  author={de Sterke, C Martijn and Blanco-Redondo, Andrea},
  journal={Optics Communications},
  volume={541},
  pages={129560},
  year={2023},
  publisher={Elsevier}
}

@article{runge2021infinite,
  title={Infinite hierarchy of solitons: Interaction of Kerr nonlinearity with even orders of dispersion},
  author={Runge, Antoine FJ and Qiang, Y Long and Alexander, Tristram J and Rafat, MZ and Hudson, Darren D and Blanco-Redondo, Andrea and de Sterke, C Martijn},
  journal={Physical Review Research},
  volume={3},
  number={1},
  pages={013166},
  year={2021},
  publisher={APS}
}

@article{sardelis2024pure,
  title={Pure-quartic solitons with PT-symmetric nonlinearity},
  author={Sardelis, Savvas and Roy, Shuva and Roy, Mrinmoy and Musslimani, Ziad and Blanco-Redondo, Andrea},
  journal={Optics Letters},
  volume={49},
  number={21},
  pages={6069--6072},
  year={2024},
  publisher={Optica Publishing Group}
}

@article{liu2022dynamic,
  title={The dynamic characteristics of pure-quartic solitons and soliton molecules},
  author={Liu, Xiaoyan and Zhang, Hongxin and Liu, Wenjun},
  journal={Applied Mathematical Modelling},
  volume={102},
  pages={305--312},
  year={2022},
  publisher={Elsevier}
}

@article{qian2022dissipative,
  title={Dissipative pure-quartic soliton fiber laser},
  author={Qian, Zi-Chen and Liu, Meng and Luo, Ai-Ping and Luo, Zhi-Chao and Xu, Wen-Cheng},
  journal={Optics Express},
  volume={30},
  number={12},
  pages={22066--22073},
  year={2022},
  publisher={Optica Publishing Group}
}

@article{baqer2025shallow,
  title={On Shallow Water Non-convex Dispersive Hydrodynamics: The Extended KdV Model: S. Baqer et al.},
  author={Baqer, Saleh and Horikis, Theodoros P and Frantzeskakis, Dimitrios J},
  journal={Water Waves},
  volume={7},
  number={2},
  pages={225--262},
  year={2025},
  publisher={Springer}
}

@article{mossman2020stability,
  title={Stability in turbulence: The interplay between shocks and vorticity in a superfluid with higher-order dispersion},
  author={Mossman, Maren E and Delikatny, Edward S and Forbes, Michael McNeil and Engels, Peter},
  journal={Physical Review A},
  volume={102},
  number={5},
  pages={053310},
  year={2020},
  publisher={APS}
}

@article{sprenger2017shock,
  title={Shock waves in dispersive hydrodynamics with nonconvex dispersion},
  author={Sprenger, Patrick and Hoefer, Mark A},
  journal={SIAM Journal on Applied Mathematics},
  volume={77},
  number={1},
  pages={26--50},
  year={2017},
  publisher={SIAM}
}

@article{conforti2013dispersive,
  title={Dispersive wave emission from wave breaking},
  author={Conforti, Matteo and Trillo, Stefano},
  journal={Optics Letters},
  volume={38},
  number={19},
  pages={3815--3818},
  year={2013},
  publisher={Optical Society of America}
}

@inproceedings{xu2012supercontinuum,
  title={supercontinuum generation in photonic crystal fiber with all-normal group velocity dispersion},
  author={Xu, Yongzhao and Zhang, Xia and Hou, Shanglin and Liu, Minxia and Wang, Hongcheng and Ling, Dongxiong},
  booktitle={2012 8th International Conference on Wireless Communications, Networking and Mobile Computing},
  pages={1--4},
  year={2012},
  organization={IEEE}
}

@article{hooper2011coherent,
  title={Coherent supercontinuum generation in photonic crystal fiber with all-normal group velocity dispersion},
  author={Hooper, Lucy E and Mosley, Peter James and Muir, Alistair C and Wadsworth, William J and Knight, Jonathan C},
  journal={Optics Express},
  volume={19},
  number={6},
  pages={4902--4907},
  year={2011},
  publisher={Optical Society of America}
}

@article{el2016radiating,
  title={Radiating dispersive shock waves in non-local optical media},
  author={El Gennady, A and Smyth, Noel F},
  journal={Proceedings. Mathematical, Physical, and Engineering Sciences/The Royal Society},
  volume={472},
  number={2187},
  pages={20150633},
  year={2016}
}

@article{conforti2014resonant,
  title={Resonant radiation shed by dispersive shock waves},
  author={Conforti, Matteo and Baronio, Fabio and Trillo, Stefano},
  journal={Physical Review A},
  volume={89},
  number={1},
  pages={013807},
  year={2014},
  publisher={APS}
}

@article{el2006unsteady,
  title={Unsteady undular bores in fully nonlinear shallow-water theory},
  author={El, GA and Grimshaw, Roger HJ and Smyth, Noel F},
  journal={Physics of Fluids},
  volume={18},
  number={2},
  year={2006},
  publisher={AIP Publishing}
}

@article{trillo2017wave,
  title={Wave-Breaking and Dispersive Shock Wave Phenomena in Optical Fibers},
  author={Trillo, Stefano and Conforti, Matteo},
  journal={Shaping Light in Nonlinear Optical Fibers},
  pages={325--349},
  year={2017},
  publisher={Wiley Online Library}
}

@article{hoefer2006dispersive,
  title     = {Dispersive and Classical Shock Waves in Bose--Einstein Condensates and Gas Dynamics},
  author    = {Hoefer, Mark A. and Ablowitz, M. J. and Coddington, I. and Cornell, Eric A. and Engels, P. and Schweikhard, V.},
  journal   = {Physical Review A},
  volume    = {74},
  number    = {2},
  pages     = {023623},
  year      = {2006}
}

@article{chang2008formation,
  title     = {Formation of Dispersive Shock Waves by Merging and Splitting Bose--Einstein Condensates},
  author    = {Chang, Jia J. and Engels, Peter and Hoefer, Mark A.},
  journal   = {Physical Review Letters},
  volume    = {101},
  number    = {17},
  pages     = {170404},
  year      = {2008}
}

@book{kevrekidis2008emergent,
  title     = {Emergent Nonlinear Phenomena in Bose--Einstein Condensates},
  author    = {Kevrekidis, Panayotis G. and Frantzeskakis, Dimitri J. and Carretero-Gonz{\'a}lez, Ricardo},
  year      = {2008},
  publisher = {Springer}

}

\end{document}